\documentclass[a4paper,11pt,DIV=12,abstract=true,numbers=noenddot,titlepage=false,twocolumn=false,draft=false]{scrartcl}
\pdfoutput=1

\usepackage[usenames,dvipsnames]{color}
  \definecolor{hgreen}{rgb}{0,.3,0}
  \definecolor{hred}{rgb}{.3,0,0}
  \definecolor{hblue}{rgb}{0,0,.3}
  \definecolor{LightGray}{gray}{0.95}
  \definecolor{gray}{gray}{0.6}
\usepackage[
	    colorlinks=true,
	    linkcolor=hblue,
	    citecolor=hgreen,
	    filecolor=hblue,
	    urlcolor=hred
	    ]{hyperref}
\usepackage{mathrsfs}
\usepackage[intlimits]{amsmath}
\usepackage{amssymb}
\usepackage{slashed}
\usepackage[titletoc,title]{appendix}
\usepackage[affil-it]{authblk}
\usepackage[numbers,sort&compress]{natbib}
\usepackage{graphicx}
\usepackage{physics}
\usepackage{bbm}
\usepackage{bbold}

\setcapindent{1em}
\setkomafont{captionlabel}{\bfseries}
\setkomafont{caption}{\itshape}

\definecolor{Blu}{rgb}{0.,0.,1.}
\definecolor{Red}{rgb}{1.,0.,0.}
\definecolor{Green}{rgb}{0.,1.,0.}
\definecolor{Purple}{rgb}{0.5,0.,0.5}

\newcommand{\rS}{\text{S}}
\newcommand{\GF}{G_F}

\begin{document}
\renewcommand\Authands{, }

\title{\boldmath 
Electroweak Corrections to the Charm-Top-Quark Contribution to $\epsilon_K$
}

\date{\today}
\author{Joachim~Brod%
        \thanks{\texttt{joachim.brod@uc.edu}}}
\author{Sandra~Kvedarait\.e%
        \thanks{\texttt{kvedarsa@ucmail.uc.edu}}}
\author{Zachary~Polonsky%
        \thanks{\texttt{polonsza@mail.uc.edu}}}
\author{Ahmed~Youssef%
        \thanks{\texttt{youssead@ucmail.uc.edu}}}
	\affil{{\large Department of Physics, University of Cincinnati, Cincinnati, OH 45221, USA}}

\maketitle

\begin{abstract}
  We calculate the leading-logarithmic and next-to-leading-logarithmic
  electroweak corrections to the charm-top-quark contribution to the
  effective $|\Delta S| = 2$ Lagrangian, relevant for the parameter
  $\epsilon_K$. We find that these corrections lead to a $-0.5\%$
  shift in the corresponding Wilson coefficient. Moreover, our
  calculation removes an implicit ambiguity in the standard-model
  prediction of $\epsilon_K$, by fixing the renormalization scheme of
  the electroweak input parameters.
\end{abstract}
\setcounter{page}{1}

\section{Introduction\label{sec:introduction}}

Indirect CP violation in the neutral kaon system, parameterized by
$\epsilon_K$, is one of the most sensitive precision probes of new
physics. The parameter $\epsilon_K$ can be expressed to excellent
approximation as~\cite{Anikeev:2001rk}
\begin{equation}\label{eq:ek:def}
  \epsilon_K \equiv e^{i\phi_\epsilon} \sin\phi_\epsilon \frac{1}{2}
  \arg \bigg( \frac{-M_{12}}{\Gamma_{12}} \bigg)\,.
\end{equation}
Here, $\phi_\epsilon = \arctan(2\Delta M_K/\Delta\Gamma_K)$, with
$\Delta M_K$ and $\Delta\Gamma_K$ the mass and lifetime difference of
the weak eigenstates $K_L$ and $K_S$. $M_{12}$ and $\Gamma_{12}$ are
the Hermitian and anti-Hermitian parts of the Hamiltonian that
determines the time evolution of the neutral kaon system. The
short-distance contributions to $\epsilon_K$ are then contained in the
matrix element $M_{12} = - \langle K^0 | \mathcal{L}^{\Delta\rS =
  2}_{f=3}| \bar K^0 \rangle / (2\Delta M_K)$, up to higher powers in
the operator-product expansion.

Experimentally,
$|\epsilon_K| = (2.228 \pm 0.0011) \times
10^{-3}$~\cite{Workman:2022ynf}, with an uncertainty at the permil
level. From the theory side, recent progress indicates that we will be
able to predict $\epsilon_K$ in the Standard Model (SM) with an
uncertainty at the percent level in the not-so-far future. Currently,
the combined perturbative uncertainty is of the order of $3\%$, while
the non-perturbative uncertainty is of the order of
$3.5\%$~\cite{Brod:2019rzc}. Interestingly, both these errors can in
principle be reduced by perturbative calculations, the first by
computing the three-loop QCD corrections to the top-quark contribution
to $\epsilon_K$, and the second by computing the two-loop conversion
to the $\overline{\text{MS}}$ scheme of the hadronic matrix element.
The non-local long-distance contributions to $\epsilon_K$, estimated
in Refs.~\cite{Proceedings:2001rdi} and~\cite{Buras:2010pza}, can be
improved in the future with lattice calculations (see
Ref.~\cite{RBC:2020kdj} for recent results).

With theory uncertainties approaching the percent level, also
parametrically smaller corrections have been taken into consideration
recently. The power corrections to the effective
Lagrangian~\cite{Cata:2003mn} have been revisited in an extended
analysis~\cite{Ciuchini:2021zgf}, leading to a one-percent increase of
the SM prediction of $\epsilon_K$. On the perturbative side, the
electroweak corrections to the top-quark contribution to $\epsilon_K$
have been calculated by some of us~\cite{Brod:2021qvc}.

In this work, we complete the analysis of the leading perturbative
electroweak and QED corrections to $\epsilon_K$ by considering the
mixed charm-top contributions. The paper is organized as follows. The
analytic results, including the details of our calculation, are
presented in Sec.~\ref{sec:ew}. The numerical evaluation, as well as a
discussion of the results, can be found in Sec.~\ref{sec:conclusion},
where we also give an updated SM prediction for
$\epsilon_K$. App.~\ref{sec:scheme} contains the definition of
evanescent operators used in our calculation.

\section{Electroweak corrections in the charm-top sector}\label{sec:ew}

In this section we provide the details of the renormalization-group
(RG) analysis. We will show the analytic results only for the
electroweak and QED corrections; the QCD corrections to
$\tilde C_{S2}^{ut}$ have already been presented in
Refs.~\cite{Herrlich:1993yv, Brod:2010mj} and can be transcribed to
our convention for the effective Lagrangian as explained in
Ref.~\cite{Brod:2019rzc}. In particular, the NNLL QCD results are not
needed as an ingredient of our calculation. Of course, they are
included in our final numerics.

\subsection{The effective Lagrangians}

As shown in Ref.~\cite{Brod:2019rzc}, it is advantageous to choose the
effective Lagrangian describing the $|\Delta S| = 2$ transition in the
three-flavor theory as
\begin{equation}\label{eq:LS2}
\begin{split}
	\mathcal{L}^{|\Delta S|=2}_{f=3} = - \frac{\GF^2 M_W^2}{4 \pi^2}
    \big[ \lambda_u^2 \tilde C_{S2}^{uu}(\mu) + \lambda_t^2 \tilde C_{S2}^{tt}
    (\mu) + \lambda_u \lambda_t \tilde C_{S2}^{ut}(\mu) \big] \tilde Q_{S2}
      + \textrm{h.c.} + \dots \,,
\end{split}
\end{equation}
because then the higher-order QCD corrections are small. Here, $G_F$
denotes Fermi's constant, and $M_W$ the $W$-boson mass. The parameters
$\lambda_i \equiv V_{is}^* V_{id}^{\phantom{*}}$ comprise the
Cabibbo-Kobayashi-Maskawa (CKM) matrix elements. The $|\Delta S| = 2$
transition is induced by the local operator
\begin{equation}\label{eq:QS2}
  \tilde Q_{S2} = (\bar s_L \gamma_\mu d_L)(\bar s_L \gamma_\mu d_L) \,,
\end{equation}
where $s_L$, $d_L$ denote the left-handed strange- and down-quark
fields, respectively. The ellipsis in Eq.~\eqref{eq:LS2} denotes
non-local contributions as well as the contribution of higher
dimension operators~\cite{Cata:2003mn, Ciuchini:2021zgf}. In the PDG
phase convention for the CKM matrix, $\lambda_u$ is real and the
coefficient $\tilde C_{S2}^{uu}$ does not affect $\epsilon_K$ (it does
contribute to the kaon mass difference $\Delta M_K$). The coefficient
$\tilde C_{S2}^{tt}$ depends on the top-quark mass and is independent
of the charm-quark mass to excellent approximation. It is known
including next-to-leading-logarithmic (NLL) QCD
corrections~\cite{Buras:1990fn}, while the electroweak corrections
have been presented in Ref.~\cite{Brod:2021qvc}. The coefficient
$\tilde C_{S2}^{ut}$, on the other hand, depends on both the charm and
top masses and has been predicted including
next-to-next-to-leading-logarithmic (NNLL) QCD
corrections~\cite{Herrlich:1993yv, Brod:2010mj, Brod:2019rzc}. Here,
we calculate the electroweak corrections to $\tilde C_{S2}^{ut}$.

The Lagrangian~\eqref{eq:LS2} is valid below the charm-quark
scale. Its Wilson coefficients are obtained by matching from the
effective four- and five flavor Lagrangians
\begin{equation}\label{eq:lag:s2:ut}
\begin{split}
  \mathcal{L}_{f=4,5}^\text{eff}
& = - \frac{4 \GF}{\sqrt{2}}
    \Bigg[ \sum_{q,q'=u,c} \!\!\! V_{qs}^\ast V_{q'd} (C_{+} Q_+^{qq'} + C_{-} Q_-^{qq'})
           - \lambda_t \sum_{i=3,6} C_i Q_i \Bigg] \\
& \quad - \frac{\GF^2 M_W^2}{4\pi^2} \lambda_t^2 C_{S2}^{tt} Q_{S2}
- 8 \GF^2 \big( \lambda_u \lambda_t + \lambda_t^2 \big) \tilde{C}_7 \tilde{Q}_{7}
       + \text{h.c.} \,,
\end{split}
\end{equation}
after the appropriate RG evolution, as described below. The first line
in Eq.~\eqref{eq:lag:s2:ut} contains the $|\Delta S|=1$
current-current operators, defined as
\begin{equation}
  Q_\pm^{qq'}
= \frac{1}{2}
  \big(     (\bar s_L^\alpha \gamma_\mu q_L^\alpha) (\bar q_L^{\prime\beta} \gamma^\mu d_L^\beta)
        \pm (\bar s_L^\alpha \gamma_\mu q_L^\beta) (\bar q_L^{\prime\beta} \gamma^\mu d_L^\alpha) \big) \,.
\end{equation}
Here, $\alpha$, $\beta$ are $SU(3)$ color indices. The QCD-penguin
operators $Q_i$, $i = 3,\ldots,6$, are defined, e.g., in
Ref.~\cite{Brod:2010mj}. They are neglected in this work as they
constitute a percent-level correction to our numerically small results
(see Sec.~\ref{sec:conclusion} below; cf. also
Ref.~\cite{Herrlich:1996vf}). The $|\Delta S|=1$ operators mix, via
bilocal insertions, into the local $|\Delta S|=2$ operator. For the
contributions proportional to $\lambda_u \lambda_t$, the
Glashow-Iliopoulos-Maiani mechanism ensures that the mixing starts at
order $m_c^2$; it is therefore convenient to define a rescaled version
of the $Q_{S2}$ operator as
\begin{equation}\label{eq:def:Qt7}
  \tilde{Q}_{7} = \frac{m_c^2}{g_s^2\mu^{2\epsilon}} (\bar s_L \gamma_\mu d_L)(\bar s_L \gamma_\mu d_L) \,.
\end{equation}
This operator is formally of dimension eight. The appearance of the
strong coupling constant in the denominator takes account of the large
logarithm in the LO result.

\subsection{RG Analysis of the charm-top contribution}

To begin, we briefly discuss the structure of the RG-improved
perturbation series. Recall that the leading QCD RG evolution of
$\tilde C_{S2}^{ut}$ reproduces the large logarithm
$\log (m_c^2/M_W^2)$ that appears in the (fixed-order) Inami-Lim
function~\cite{Inami:1980fz}, and sums this logarithm to all
orders. As these leading-order boxes involve no gluon exchange, it is
conventional to rescale the $|\Delta S| = 2$ effective
operator~\eqref{eq:QS2} with an inverse power of $g_s^2$ (see
Eq.~\eqref{eq:def:Qt7}). In this way, the leading-logarithmic (LL)
series has the standard form, with terms proportional to
$(\alpha_s \log)^n$, where $n = 1,\ldots$. The terms in the NLL and
NNLL series are then proportional to $\alpha_s (\alpha_s \log)^n$ and
$\alpha_s^2 (\alpha_s \log)^n$, respectively.

Here, we will sum the two series whose terms are proportional to
$\alpha \alpha_s^{n} \log^{n+1}$ (``LL QED'') and
$\alpha (\alpha_s \log)^n$ (``NLL QED''). The former series receives
contributions only from the one-loop QED running of the
current-current operators (see below), while the latter requires the
calculation of the one-loop electroweak initial conditions in the
current-current sector, the mixed QED-QCD RG evolution of the
current-current operators, and the QED corrections to the anomalous
dimension tensor, encoding the mixing of current-current operators
into the $|\Delta S| = 2$ sector. In contrast to the case of
$C_{S2}^{ut}(\mu)$, the two-loop electroweak initial condition of the
Wilson coefficient in the $|\Delta S| = 2$ sector is not needed, due
to the presence of the large logarithm in the leading-order (LO)
result. In addition to summing these series, our calculation fixes the
renormalization scheme of the electroweak input parameters. That is
achieved by normalizing the initial conditions of the current-current
Wilson coefficients to the muon decay
constant~\cite{Sirlin:1981ie}. In this way, a large part of the
radiative corrections is absorbed into the measured value of the muon
decay rate~\cite{Brod:2008ss}, and $G_F$ is the only requisite
electroweak input parameter for our calculation.


The actual RG analysis involves the determination of the initial
conditions of the Wilson coefficients at the electroweak scale, and
the subsequent RG evolution down to the hadronic scale where the local
hadronic matrix element is evaluated. The steps of this analysis have
been discussed extensively in the literature and no new conceptual
questions arise in our analysis, so we can afford to be brief in our
exposition.

Expanding all Wilson coefficients in the five- and four-flavor
effective theories as
\begin{equation}
  C_i(\mu)
=   C_i^{(0)}(\mu)
  + \frac{\alpha}{\alpha_s(\mu)} C_i^{(e)}(\mu)
  + \frac{\alpha}{4\pi} C_i^{(es)}(\mu)\,,
\end{equation}
we find by performing an explicit matching calculation at the
electroweak scale
\begin{equation}
	\begin{split}
		C_\pm^{(0)}(\mu_W) &= 1\,, \quad
  		C_\pm^{(e)}(\mu_W) = 0\,, \quad
		C_\pm^{(es)}(\mu_W) = - \frac{22}{9}
			- \frac{4}{3} \log\frac{\mu_W^2}{M_Z^2}\,,
	\end{split}
\end{equation}
consistent with the results in the literature~\cite{Gambino:2000fz,
  Gambino:2001au}. The initial conditions for the Wilson coefficients
in the $|\Delta S|=2$ sector vanish at this order, i.e., we have
$\tilde C_7^{(e)}(\mu_W) = \tilde C_7^{(es)}(\mu_W) = 0$.

\begin{figure}[t]
  \centering
  \includegraphics[width=0.6\textwidth]{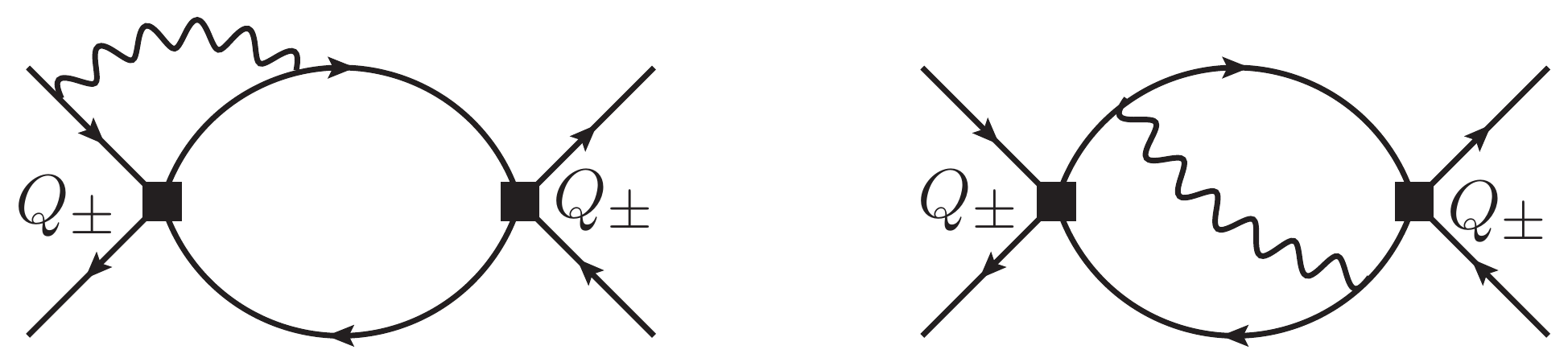}
  \caption{Sample Feynman diagrams with bilocal insertions of the
    current-current operators $Q_\pm^{qq'}$. Solid lines denote
    appropriate quark flavors, and wavy lines denote photons.
    \label{fig:feynman}}
\end{figure}

In order to evolve the Wilson coefficients down to the hadronic scale,
we need to solve the set of RG equations
\begin{equation}\label{eq:rge:single:in}
  \mu \frac{d}{d\mu} C_{i}(\mu)
= C_{j}(\mu) \gamma_{ji} \,, \quad i,j = +,- \,,
\end{equation}
and
\begin{equation}\label{eq:rge:double:in}
  \mu \frac{d}{d\mu} \tilde{C}_{7}(\mu)
= \tilde{C}_{7}(\mu) \tilde{\gamma}_{77} 
  + \sum_{k,l=+,-} C_{k}(\mu) C_{l}(\mu) \hat\gamma_{kl,7} \,.
\end{equation}
Here,
$\tilde{\gamma}_{77} = \tilde\gamma_{S2} + 2 \gamma_{m} + 2 \beta$ is
given in terms of the anomalous dimension $\tilde\gamma_{S2}$ of the
local operator $\tilde Q_{S2}$. The quark anomalous dimension and the
beta function appear because of the explicit factors of $m_c$ and
$g_s$ in the definition of $\tilde Q_7$. Further, $\gamma_{ij}$
denotes the anomalous dimension matrix in the current-current sector,
and $\hat\gamma_{kn,7}$ is the anomalous dimension tensor, describing
the mixing of the dimension-six operators into $\tilde Q_7$. Defining
$dg_s/d\log\mu = \beta$, with
\begin{equation}
\beta(g_s,e) =
-\beta_0 \frac{g_s^3}{16\pi^2}
-\beta_1 \frac{g_s^5}{(16\pi^2)^2}
-\beta_{es} \frac{e^2g_s^3}{(16\pi^2)^2}
+ \ldots \,,
\end{equation}
and $dm/d\log\mu = - m\gamma_m$, with
\begin{equation}\label{eq:gammam:exp}
 \gamma_m(g_s, e) =
 \gamma_m^{(0)} \frac{g_s^2}{16\pi^2}
+\gamma_m^{(1)} \frac{g_s^4}{(16\pi^2)^2}
+\gamma_m^{(e)} \frac{e^2}{16\pi^2}
+\gamma_m^{(es)} \frac{e^2g_s^2}{(16\pi^2)^2}
+ \ldots \,,
\end{equation}
we have~\cite{Buchalla:1995vs, Brod:2021qvc}
\begin{equation}\label{eq:gammat}
\tilde\gamma_{S2}^{(0)} =
4\,,
\quad
\tilde\gamma_{S2}^{(e)} =
\frac{4}{3}\,,
\quad
\tilde\gamma_{S2}^{(es)} =
-\frac{148}{9}\,,
\end{equation}
and
\begin{equation}\label{eq:gammaexp}
\gamma_m^{(0)} = 8 \,,
\qquad
\gamma_m^{(e)} = \frac{8}{3} \,,
\qquad
\gamma_m^{(es)} = \frac{32}{9} \,,
\end{equation}
\begin{equation}\label{eq:betaqcdexp}
\beta_{0} = 11-\frac{2}{3}f \,,
\qquad
\beta_{e} = 0 \, ,
\qquad
\beta_{es} = -\frac{8}{9}(f_u+\frac{f_d}{4}) \,,
\end{equation}
where $f_u$ and $f_d$ denote the number of up- and down-type quark
flavours, and $f = f_u + f_d$. Moreover~\cite{Buras:1992zv}
\begin{equation}
\gamma^{(0)} =
\begin{pmatrix}
  4&0\\0&-8
\end{pmatrix}\,,
\quad
\gamma^{(e)} =
\begin{pmatrix}
  -\frac{8}{3}&0\\0&-\frac{8}{3}
\end{pmatrix}\,,
\quad
\gamma^{(es)} =
\begin{pmatrix}
  \frac{107}{9}&-18\\[1ex]
  -9&\frac{38}{9}
\end{pmatrix}\,,
\end{equation}
and
\begin{equation}
\hat \gamma_{kl,7}^{(0)} =
\begin{pmatrix}
  3&-1\\-1&1
\end{pmatrix}\,,
\quad
\hat \gamma_{kl,7}^{(e)} =
\begin{pmatrix}
  0&0\\0&0
\end{pmatrix}\,,
\quad
\hat \gamma_{kl,7}^{(es)} =
\begin{pmatrix}
  43&-\frac{43}{3}\\[1ex]
  -\frac{43}{3}&\frac{43}{3}
\end{pmatrix}\,,
\end{equation}
with an expansion defined in analogy to Eq.~\eqref{eq:gammam:exp}.
The result for $\hat \gamma_{kl,7}^{(es)}$ is new. It has been
calculated in terms of the renormalization
constants~\cite{Brod:2009ezt} for bilocal insertions of
current-current operators (see Fig.~\ref{fig:feynman} for sample
Feynman diagrams). All diagrams have been calculated using
self-written \texttt{FORM}~\cite{Vermaseren:2000nd} routines,
implementing the two-loop recursion presented in
Refs.~\cite{Davydychev:1992mt, Bobeth:1999mk}. The amplitudes were
generated using \texttt{qgraf}~\cite{Nogueira:1991ex}. We used the
algorithm in Ref.~\cite{Chetyrkin:1997fm} to isolate the UV
divergences.

Solving the inhomogeneous system of differential
equations~\eqref{eq:rge:single:in} and~\eqref{eq:rge:double:in} is
tedious. It is, however, straightforward to verify that
Eqs.~\eqref{eq:rge:single:in} and~\eqref{eq:rge:double:in} are
equivalent to the homogeneous system of equations
\begin{equation}\label{eq:rge:combo}
  \mu \frac{d}{d\mu} D(\mu) = D(\mu) \gamma \,,
\end{equation}
with\footnote{All penguin contributions to the QCD RG evolution are
  included in our final numerics, and have been evaluated using a
  straightforward generalization of these definitions. We checked
  explicitly that we reproduce the QCD results in the literature, up
  to NNLL.}
\begin{equation}\label{eq:rge:combo:C}
  D(\mu) = 
  \begin{pmatrix}
    C_{+}(\mu)^2 \\ C_{+}(\mu) C_{-}(\mu) \\ C_{-}(\mu)^2 \\ \tilde{C}_{7}
  \end{pmatrix}
\end{equation}
and
\begin{equation}\label{eq:rge:combo:ADM}
  \gamma = 
  \begin{pmatrix}
    2 \gamma_{++} & \gamma_{+-} & 0 & \hat \gamma_{++,7} \\
    2 \gamma_{-+} & \gamma_{++} + \gamma_{--} & 2 \gamma_{+-} & \hat \gamma_{+-,7} + \hat \gamma_{-+,7} \\
    0 & \gamma_{-+} & 2 \gamma_{--} & \hat \gamma_{--,7} \\
    0 & 0 & 0 & \tilde \gamma_{77}
  \end{pmatrix} \,.
\end{equation}
This can be solved using standard methods (see, for instance,
Reference~\cite{Gorbahn:2004my}). The RG evolution can be conveniently
written in terms of an evolution matrix, such that $D(\mu) = D(\mu_0)
U(\mu_0,\mu,\alpha)$. We expand
\begin{equation}\label{eq:U:expansion}
\begin{split}
U(\mu_0,\mu,\alpha) = 
   U^{(0)}(\mu_0,\mu)
 + \frac{\alpha}{\alpha_s(\mu)}U^{(e)}(\mu_0,\mu)
 + \frac{\alpha}{4\pi}U^{(se)}(\mu_0,\mu)
 + \ldots \,.
\end{split}
\end{equation}
Here, $\mu_0$ and $\mu$ denote the generic ``high'' and ``low'' scale
of the RG evolution (i.e., $\mu_0 = \mu_t$ and $\mu = \mu_b$ in the
five-flavor, and $\mu_0 = \mu_b$ and $\mu = \mu_c$ in the four-flavor
theory). We find the following contributions to the Wilson coefficient
at the low scale:
\begin{align}\label{eq:C:low:coefficients}
D^{(0)}(\mu) & = D^{(0)}(\mu_0) U^{(0)}(\mu_0,\mu)\,,\\
D^{(e)}(\mu) & = D^{(0)}(\mu_0) U^{(e)}(\mu_0,\mu) + \eta^{-1} D^{(e)}(\mu_0) U^{(0)}(\mu_0,\mu)\,,\\
\begin{split}
D^{(se)}(\mu) & = \eta D^{(1)}(\mu_0) U^{(e)}(\mu_0,\mu)
                 + \eta^{-1} D^{(e)}(\mu_0) U^{(1)}(\mu_0,\mu) \\
            & \quad + D^{(se)}(\mu_0) U^{(0)}(\mu_0,\mu)
                    + D^{(0)}(\mu_0) U^{(se)}(\mu_0,\mu)\,,
\end{split}
\end{align}
where we have introduced the ratio
$\eta=\alpha_s(\mu_0)/\alpha_s(\mu)$. The explicit expression for the
evolution matrix, in terms of the anomalous dimensions of the Wilson
coefficients, can be found in Ref.~\cite{Buras:1993dy}.

At the bottom threshold, $\mu_b \sim m_b$, the bottom quark is removed
as a dynamical degree of freedom. Numerically, the impact of this
threshold correction is small. In fact, since we neglect the
contribution of penguin operators for the QED and electroweak
corrections, the only effect is the decoupling of $\alpha_s$ from
$f=5$ to $f=4$:
\begin{equation}
\alpha_s^{(5)} = \alpha_s^{(4)}
\bigg( 1 + \frac{2}{3} \frac{\alpha_s^{(4)}}{4\pi} \log(\mu_b^2/m_b(\mu_b)^2)\bigg) \,,
\end{equation}
leading to an additional logarithmic contribution to all Wilson
coefficients. Requiring the equality of all Green's functions at the
matching scale and writing
$\delta C(\mu_b) = C_{f=5}(\mu_b) - C_{f=4}(\mu_b)$ we find, for the
dimension-six Wilson coefficients,
\begin{equation}
  \delta C_i^{(0)} = 0 \,, \quad \delta C_i^{(e)} = 0 \,, \quad
  \delta C_i^{(es)} = \frac{2}{3} C_i^{(e)} \log \bigg(\frac{\mu_b^2}{m_b(\mu_b)^2} \bigg) \,,
\end{equation}
and for the dimension-eight Wilson coefficient (taking into account
the factor $m_c^2/g_s^2$ in the definition of the operator)
\begin{equation}
  \delta \tilde C_7^{(0)} = 0 \,, \quad \delta \tilde C_7^{(e)} = 0 \,, \quad
  \delta \tilde C_7^{(es)} = \frac{4}{3} \tilde C_7^{(e)} \log \bigg(\frac{\mu_b^2}{m_b(\mu_b)^2} \bigg) \,.
\end{equation}

At the scale $\mu_c \sim m_c$ the charm quark is removed from the
theory as a dynamical degree of freedom, and the effective Lagrangian
is now given by Eq.~\eqref{eq:LS2}. Requiring the equality of the
Green's functions in both the four-flavor and three-flavor theories at
the charm-quark scale leads to the matching condition
\begin{equation}
  \sum_{k,l=+,-} C_k(\mu_c) C_l(\mu_c) \langle Q_{k} Q_l 
  \rangle(\mu_c) + {\tilde C}_7(\mu_c) \langle {\tilde Q}_7
  \rangle(\mu_c)
= \frac{M_W^2}{32\pi^2} {\tilde C}_{S2}^{ut}(\mu_c) 
  \langle {\tilde Q}_{S2} \rangle(\mu_c) \,, 
\end{equation}
where the angle brackets denote the partonic $|\Delta S| = 2$ matrix
elements. We parameterize these matrix elements in the following way:
\begin{equation}
  \langle {\tilde Q}_7 \rangle
    = r_{S2} \langle {\tilde Q}_{7} \rangle^{(0)} \,, \quad
  \langle {\tilde Q}_{S2} \rangle
    = r_{S2} \langle {\tilde Q}_{S2} \rangle^{(0)} \,, \quad
  \text{and} \,\,
  \langle Q_i Q_j \rangle (\mu_c)
    = \frac{m_c^2(\mu_c)}{32\pi^2M_W^2} r_{ij,S2}^{ut} \langle {\tilde Q}_{S2} \rangle^{(0)} \,. 
\end{equation}
Taking into account the explicit factor of $m_c^2/g_s^2$ in the
definition of $\tilde Q_7$, we expand the Wilson coefficient
$\tilde{C}_{S2}^{ut}$ as
\begin{equation}
    \tilde{C}_{S2}^{ut}
=   \frac{4\pi}{\alpha_s^{(3)}} \tilde{C}_{S2}^{ut(0)}
  + \frac{4\pi\alpha}{(\alpha_s^{(3)})^2} \tilde{C}_{S2}^{ut(e)}
  + \frac{\alpha}{\alpha_s^{(3)}} \tilde{C}_{S2}^{ut(es)}
  + \ldots \,,
\end{equation}
and find the following contributions to the matching:
\begin{align}
\tilde{C}_{S2}^{ut(0)}(\mu_c)
  & = 2 \frac{m_c^2(\mu_c)}{M_W^2}
      \tilde{C}_{7}^{(0)}(\mu_c) \,, \label{eq:matchmc0} \\
\tilde{C}_{S2}^{ut(e)}(\mu_c)
  & = 2 \frac{m_c^2(\mu_c)}{M_W^2}
      \tilde{C}_{7}^{(e)}(\mu_c) \,, \label{eq:matchmce} \\
\tilde{C}_{S2}^{ut(es)}(\mu_c)
  & = 2 \frac{m_c^2(\mu_c)}{M_W^2}
      \bigg[ \tilde{C}_{7}^{(es)}(\mu_c)
             - \frac{4}{3} \tilde{C}_{7}^{(e)}(\mu_c) \log \frac{\mu_c^2}{m_c(\mu_c)^2}
      \bigg] \notag \\
  & \quad + \frac{m_c^2(\mu_c)}{M_W^2}
    \Big[ C_i^{(0)}(\mu_c) C_j^{(e)}(\mu_c) + C_i^{(e)}(\mu_c) C_j^{(0)}(\mu_c) \Big]
    r_{ij,S2}^{ut,(0)} \,. \label{eq:matchmces}
\end{align}
Here, $m_c(\mu_c)$ denotes the running charm-quark mass, including the
leading QED running. We see that for the electroweak corrections, the
LO matching result is sufficient. According to
Ref.~\cite{Brod:2019rzc}, it can be taken as
$r_{ij,S2}^{ut} = 2r_{ij,S2}^{cc} - r_{ij,S2}^{ct}$ in terms of the
results in Ref.~\cite{Herrlich:1993yv, Brod:2010mj, Brod:2019rzc}. We
find
\begin{equation}
  r_{ij,S2}^{ut,(0)} =
\begin{pmatrix}
\frac{9}{2}-3\log\frac{\mu_c^2}{m_c^2(\mu_c)}&-\frac{3}{2}+\log\frac{\mu_c^2}{m_c^2(\mu_c)}\\[1ex]
-\frac{3}{2}+\log\frac{\mu_c^2}{m_c^2(\mu_c)}&\frac{3}{2}-\log\frac{\mu_c^2}{m_c^2(\mu_c)}
\end{pmatrix}\,.
\end{equation}

Finally, the RG evolution in the effective three-flavor theory
involves only the single physical operator $\tilde Q_{S2}$, with
anomalous dimension given in Eq.~\eqref{eq:gammat}. As discussed in
detail in Ref.~\cite{Brod:2021qvc}, the mixed two-loop anomalous
dimension $\tilde\gamma_{S2}^{(es)}$ is renormalization-scheme
independent, which prevents us from extending the definition of the
scheme-independent correction factors $\eta_{ut}$ to include
electroweak corrections. Instead, we have to work with Wilson
coefficients directly. In particular, our result for
$\tilde C_{S2}^{ut}$ is not independent of the renormalization scheme.

Of course, this residual scheme dependence will cancel once we
multiply $\tilde C_{S2}^{ut}$ by the hadronic matrix element of the
local operator $\tilde Q_{S2}$, evaluated including the leading QED
corrections. While this matrix element is a non-perturbative quantity,
and the QED corrections are not (yet) available, it is easy to
calculate the scheme-dependent part~\cite{Brod:2021qvc}. As a cross
check of our calculation, we kept the definition of all contributing
evanescent operators arbitrary (see App.~\ref{sec:scheme}) and
verified that all scheme dependence completely cancels in the product
of the Wilson coefficient and (the scheme-dependent part of) the
matrix element. In particular, we verified that the only left-over
dependence of $\tilde C_{S2}^{ut}$ is on the parameter $a_{11}$; all
other parameters cancel.\footnote{In this context we note that one of
  the statements made below Eq.~(C.12) in Ref.~\cite{Brod:2021qvc} is
  not correct: even for scheme-independent $\tilde\gamma_{S2}^{(es)}$,
  the corresponding evolution matrix $U^{(se)}$ does depend on the
  renormalization scheme via its dependence on the two-loop QCD
  anomalous dimension, $\tilde\gamma_{S2}^{(1)}$. However, our
  analytic check shows that this dependence drops out completely in
  the product of Wilson coefficient and the (known) QCD part of the
  hadronic matrix element. The same is true in the top-quark sector.}
We will discuss the numerical size of the scheme-dependent term in the
following section.

\section{Discussion and Conclusion}\label{sec:conclusion}

\begin{figure}[t]
  \centering
  \includegraphics[width=0.48\textwidth]{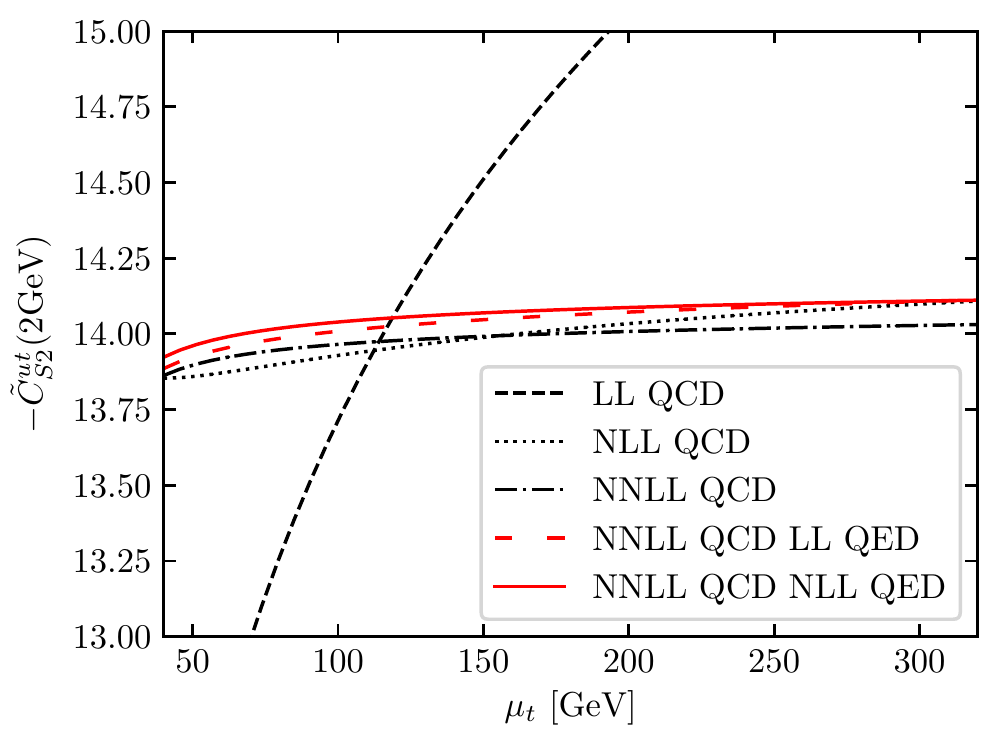}~~~
  \includegraphics[width=0.48\textwidth]{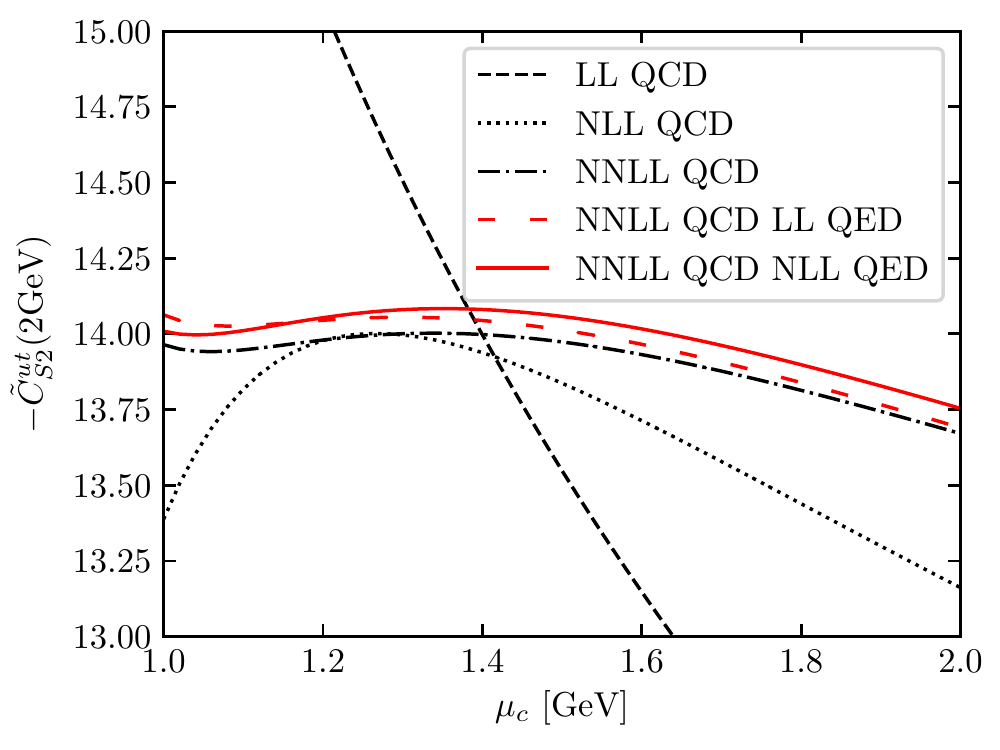}
  \caption{Residual dependence of the Wilson cofficient
    $\tilde C_{S2}^{ut}(2\,\mathrm{GeV})$ on the electroweak (left
    panel) and charm-threshold (right panel) matching scales. The
    short-dashed, dotted, and dash-dotted lines show the LL, NLL, and
    NNLL QCD results, respectively. The long-dashed and solid lines
    show the results including also the LL and NLL electroweak
    corrections.
    \label{fig:mu}}
\end{figure}

To obtain a numerical estimate of the size of the electroweak
corrections, as well as an estimate of the remaining perturbative
uncertainties, we evaluate the Wilson coefficient
$\tilde C_{S2}^{ut}(2\,\text{GeV})$, including now all known QCD
corrections, and varying the electroweak and charm-threshold matching
scales in the intervals
$40\,\text{GeV} \leq \mu_t \leq 320\,\text{GeV}$ and
$1\,\text{GeV} \leq \mu_c \leq 2\,\text{GeV}$. (The dependence on the
bottom-quark matching scale is negligible in comparison.) The
resulting residual scale variation is displayed in Fig.~\ref{fig:mu}.

To obtain a final value, we fix $\mu_t = m_t$ and take the average of
the highest and lowest value of $\tilde C_{S2}^{ut}$ in the interval
for the variation of $\mu_c$, and half the difference between the
highest and lowest values as the uncertainty. Retaining only the QCD
corrections up to NNLL, we find
$\tilde C_{S2}^{ut,\text{QCD}} = - 13.84 \pm 0.17$. Including also the
LL and NLL electroweak corrections gives
$\tilde C_{S2}^{ut} = - 13.92 \pm 0.16$. This amounts to a $-0.5\%$
shift, while the uncertainty is essentially unchanged.

What is the numerical impact of the unmatched scheme-dependent term on
this result? First, recall that the only dependence is on the
parameter $a_{11}$ (see App.~\ref{sec:scheme}); all other scheme
dependence fully cancels against the corresponding terms in the
hadronic matrix element.\footnote{In practice, the hadronic matrix
  element is only converted to the $\overline{\text{MS}}$ scheme at
  NLO, such that part of the NNLO QCD scheme dependence is not
  included. However, in the conventional formalism of the $\eta$
  correction factors, the {\em perturbative} part of the result is
  scheme independent including NNLO, as far as QCD is concerned.} Part
of the dependence on $a_{11}$ of our result arises from the dependence
on the two-loop QCD anomalous dimension; any such dependence is also
canceled by the corresponding scheme dependence of the hadronic matrix
element. The only residual dependence on $a_{11}$ is that related to
the leading QED corrections to the matrix element; numerically, it is
tiny:
\begin{equation}
  \tilde C_{S2}^{ut} (2\,\text{GeV}) = -14.075 + 0.001 a_{11} \,.
\end{equation}
We obtained this number by setting all threshold matching scales to
their respective quark masses.

While a consistent estimate of the full electroweak and QED
corrections can be obtained only once a lattice calculation (or
another systematic estimate) of the QED correction to the hadronic
matrix element becomes available, we point out that this correction is
not enhanced by a large logarithm and thus of order
$\alpha/(4\pi) \sim 10^{-4}$, the same as the residual scheme
dependence. It is expected to be numerically negligible compared to
the $-0.5\%$ shift found above. It follows that for our numerics we
can safely adopt the standard definition of evanescent operators with
$a_{11} = 4$.

To summarize, given the uncancelled (but small) residual scheme
dependence of our result, we propose a temporary prescription in
analogy to the case of the top-quark contribution~\cite{Brod:2021qvc}:
we rescale the NNLL QCD value of
$\eta_{ut} = 0.402(5)$~\cite{Brod:2019rzc} by a factor of $1.005$, to
take account of the electroweak corrections.\footnote{Recall that
  $\eta_{ut}$ is defined via
  $\tilde C_{S2}^{ut} = 2 \eta_{ut} \mathscr{S}_{ut}(x_c,x_t)$, with
  the modified Inami-Lim function~\cite{Brod:2019rzc} being negative.}
Including also the power correction presented in
Ref.~\cite{Ciuchini:2021zgf}, this leads to an updated SM prediction
of
\begin{equation}
 |\epsilon_K|
 = \big( 2.170 \pm 0.065_\text{pert.} \pm 0.076_\text{nonpert.} \pm 0.153_\text{param.} \big)
   \times 10^{-3} \,.
\end{equation}
Here, the quoted errors correspond to the residual perturbative,
non-perturbative, and parametric uncertainties, respectively; see
Ref.~\cite{Brod:2019rzc} for details. We obtained this number by
employing the phenomenological expression in
Ref.~\cite{Buchalla:1995vs}, including the long-distance corrections
presented in Refs.~\cite{Proceedings:2001rdi, Buras:2010pza}.

All parametric inputs are taken from PDG~\cite{Workman:2022ynf}. In
particular, as input for the top-quark mass we use the
$\overline{\text{MS}}$ mass $m_t(m_t) = 162.92(67)\,$GeV, obtained by
converting the pole mass $M_t = 172.5(7)\,$GeV~\cite{Workman:2022ynf}
to $\overline{\text{MS}}$ at three-loop accuracy using
RUNDEC~\cite{Chetyrkin:2000yt}.



\bigskip

In summary, we calculated the leading and next-to-leading electroweak
corrections to the charm-top contribution $\tilde C_{S2}^{ut}$ to the
effective $|\Delta S| = 2$ effective Lagrangian, using RG-improved
perturbation theory. We find a small negative shift of the Wilson
coefficient, and a corresponding small positive shift of
$\epsilon_K$. A systematic estimate of the QED corrections to the
hadronic matrix element would complete our analysis.

As consistency checks, we performed the calculation in generalized
$R_\xi$ gauge for gluons and photon and verified the gauge-parameter
independence of our results. We analytically checked that our results
are independent of all matching scales, and that the dependence on the
renormalization scheme is canceling by the corresponding scheme
dependence of the hadronic matrix element.

The aim of this work is to provide a further step in the prediction of
$\epsilon_K$ with residual theoretical uncertainty at the percent
level. Further important directions of improvement are the calculation
of the three-loop QCD corrections in the top-quark sector of the
effective Lagrangian, and the NLO scheme conversion from RI/SMOM to
$\overline{\text{MS}}$ for the hadronic matrix element of the local
$|\Delta S| = 2$ operator.

\section*{Acknowledgements}

The authors acknowledge support in part by DoE grant
de-sc0011784. J.B. thanks Martin Gorbahn and Emmanuel Stamou for
discussions.

\appendix

\section{Definition of evanescent operators}\label{sec:scheme}

In the context of dimensional regularization, evanescent operators
arise in intermediate stages of the calculation because certain
relations (such as Dirac algebra and Fierz transformations) are valid
only in four space-time dimensions. In the dimension-six sector, we
define them as
\begin{align}
  E_1^{qq'(1)} & = (\bar s_L \gamma_{\mu_1\mu_2\mu_3} T^a q_L)\otimes 
		(\bar q_L' \gamma^{\mu_1\mu_2\mu_3} T^a d_L)
		 - (16 - A_{11}\epsilon - A_{12}\epsilon^2) Q_1^{qq'} \,,\\
  E_2^{qq'(1)} & = (\bar s_L \gamma_{\mu_1\mu_2\mu_3} q_L)\otimes
		(\bar q_L' \gamma^{\mu_1\mu_2\mu_3} d_L)
		 - (16 - B_{11}\epsilon - B_{12}\epsilon^2) Q_2^{qq'} \,,\\
  E_1^{qq'(2)} & = (\bar s_L \gamma_{\mu_1\mu_2\mu_3\mu_4\mu_5} T^a q_L)\otimes
		(\bar q_L' \gamma^{\mu_1\mu_2\mu_3\mu_4\mu_5} T^a d_L)
		 - \bigg(256 - A_{21}\epsilon - A_{22}\epsilon^2\bigg) Q_1^{qq'} \,,\\
  E_2^{qq'(2)} & = (\bar s_L \gamma_{\mu_1\mu_2\mu_3\mu_4\mu_5} q_L)\otimes
		(\bar q_L' \gamma^{\mu_1\mu_2\mu_3\mu_4\mu_5} d_L)
		 - \bigg(256 - B_{21}\epsilon - B_{22}\epsilon^2\bigg) Q_2^{qq'} \,.
\end{align}
while the evanescent operators in the dimension-eight sector have been
chosen as
\begin{align}
  \tilde{E}_{F} & = \frac{m_c^2}{g^2\mu^{2\epsilon}} (\bar{s}_L^{\alpha} \gamma_{\mu}
  d_L^{\beta})\otimes (\bar{s}_L^{\beta} \gamma^{\mu} d_L^{\alpha}) - \tilde{Q}_{7} \,,\\
  \tilde{E}_{7}^{(1)} & = \frac{m_c^2}{g^2\mu^{2\epsilon}}
  (\bar{s}_L^{\alpha} \gamma_{\mu_1\mu_2\mu_3}
  d_L^{\alpha})\otimes (\bar{s}_L^{\beta} \gamma^{\mu_1\mu_2\mu_3}
	d_L^{\beta}) - (16 - a_{11} \epsilon - a_{12} \epsilon^2) \tilde{Q}_{7} \,, \\   
  \tilde{E}_{8}^{(1)} & = \frac{m_c^2}{g^2\mu^{2\epsilon}}
  (\bar{s}_L^{\alpha} \gamma_{\mu_1\mu_2\mu_3}
  d_L^{\beta})\otimes (\bar{s}_L^{\beta} \gamma^{\mu_1\mu_2\mu_3}
	d_L^{\alpha}) - (16 - b_{11} \epsilon - b_{12} \epsilon^2) (\tilde{Q}_{7} + \tilde{E}_F) \,,\\
  \tilde{E}_{7}^{(2)} & = \frac{m_c^2}{g^2\mu^{2\epsilon}}
  (\bar{s}_L^{\alpha} \gamma_{\mu_1\mu_2\mu_3\mu_4\mu_5}
  d_L^{\alpha})\otimes (\bar{s}_L^{\beta} \gamma^{\mu_1\mu_2\mu_3\mu_4\mu_5}
	d_L^{\beta}) - (256 - a_{21} \epsilon - a_{22} \epsilon^2) \tilde{Q}_{7}\, , \\ 
  \tilde{E}_{8}^{(2)} & = \frac{m_c^2}{g^2\mu^{2\epsilon}}
  (\bar{s}_L^{\alpha} \gamma_{\mu_1\mu_2\mu_3\mu_4\mu_5}
  d_L^{\beta})\otimes (\bar{s}_L^{\beta} \gamma^{\mu_1\mu_2\mu_3\mu_4\mu_5}
	d_L^{\alpha}) - (256 - b_{21} \epsilon - b_{22}
  \epsilon^2) (\tilde{Q}_{7} + \tilde{E}_F) \,.
\end{align}
Note that, to facilitate an additional check on our calculation, we
have kept the coefficients in front of the $\epsilon$ terms arbitrary.
In the conventional definition of these operators,
$A_{1i} = B_{1i} = a_{1i} = b_{1i} = 4$ where $i=1,2$,
$A_{21} = B_{21} = a_{21} = b_{21} = 224$, $a_{22} = 5712/25$,
$B_{22} = 10032/25$, and $a_{22} = b_{22} = 108\,816/325$. The
evanescent operators related to $\tilde Q_{S2}$ are defined with the
same coefficients. We have checked explicitly that the terms quadratic
in $\epsilon$ do not contribute to the two-loop anomalous
dimensions. All results quoted in the main body of the paper
correspond to the conventional definition of evanescent operators.

\addcontentsline{toc}{section}{References}
\bibliographystyle{JHEP}
\bibliography{references}

\end{document}